\documentclass{appolb}
\usepackage[dvips]{color}
\usepackage{epsf,epsfig,graphicx,graphics,amsfonts,amssymb,color}
\usepackage{ulem}

\begin{document}

\title{Stock returns versus trading volume: is the correspondence more general?}

\author{Rafa{\l} Rak$^{1}$, Stanis{\l}aw Dro\.zd\.z$^{2,3}$, Jaros{\l}aw Kwapie\'n$^{2}$, Pawe{\l} O\'swi\c ecimka$^{2}$
\address{$^1$ Faculty of Mathematics and Natural Sciences, University of Rzesz\'ow, PL--35-959 Rzesz\'ow, Poland\\
$^2$ Complex Systems Theory Department, Institute of Nuclear Physics, Polish Academy of Sciences, PL--31-342 Krak\'ow, Poland\\
$^3$ Faculty of Physics, Mathematics and Computer Science, Cracow University of Technology, PL--31-155 Krak\'ow, Poland}}

\maketitle

\begin{abstract}

This paper presents a quantitative analysis of the relationship between the stock market returns and corresponding trading volumes using high-frequency data from the Polish stock market. First, for stocks that were traded for sufficiently long period of time, we study the return and volume distributions and identify their consistency with the power-law functions. We find that, for majority of stocks, the scaling exponents of both distributions are systematically related by about a factor of 2 with the ones for the returns being larger. Second, we study the empirical price impact of trades of a given volume and find that this impact can be well described by a square-root dependence: $r(V) \sim V^{1/2}$. We conclude that the properties of data from the Polish market resemble those reported in literature concerning certain mature markets.

\end{abstract}


\section{Introduction}

Every day millions of individuals all over the world make billions of orders to buy or sell stocks according to their own investment strategies and in reaction to huge amount of arriving information. These individual decisions taken together define very complex behaviour of the financial markets~\cite{arthur1999,farmer2002,burda2003,kwapien2012,mantegna2000} and lead to such characteristics of the financial data like multifractality~\cite{fisher1997,ivanova1999,dimatteo2003,matia2003,oswiecimka2005,oswiecimka2006}, long memory, nonlinear correlations~\cite{matia2003,kwapien2005,rak2005}, the leverage effect~\cite{eisler2004,bouchaud2001}, fat tails of financial data fluctuations~\cite{liu1999,gopikrishnan1999,plerou1999,matia2002,drozdz2003,clauset2009,racz2009,mu2009,%
zhou2012,gabaix2003,plerou2004,gabaix2006,farmer2004_1,stanley2008,mu2010}, known together as the financial stylized facts.

The fat tails of pdfs mean that, for example, the nature of logarithmic price fluctuations (returns) differs from the Gaussian noise model as the former can be much larger than the latter expects. The relative magnitude of the most spectacular events observed in the financial markets is a good example of this property. Let us assume that we have a bi-variate time series of price and volume recordings collected over some period of time. In general, the returns can be defined as
\begin{equation}
r(\Delta t_k) = {R (\Delta t_k) - \langle R (\Delta t_k) \rangle \over \sigma}, \qquad R (\Delta t_k) = \ln Q(t_k+\Delta t_k) - \ln Q(t_k),
\end{equation}
where $\Delta t_k$ is length of the $k$th interval of time, $Q(t)$ is stock price, $\sigma$ is standard deviation, and $\langle \cdot \rangle$ is mean over all the intervals $k=1,...,T$. For each return $r(\Delta t_k)$, there exists an appropriate trading volume $V(\Delta t_k)$. The intervals $\Delta t_k$ can be defined in various ways: they can either cover a constant number of consecutive transactions $n_{\rm T}$, be defined by constant trading volume, or be equal to each other with $\Delta t_k = \Delta t$ for all $k$'s. The latter definition is the most common one.

It was documented in literature that the non-Gaussian pdfs of the return can be described in most cases by power-law tails of the form $p(x) \sim x^{-(1+\alpha)}$ with $\alpha>0$. Exact values of the scaling index $\alpha$ depend on the return definition and vary from market to market~\cite{racz2009,zhou2012,gopikrishnan2000,plerou2007,eisler2007} and from past to present~\cite{drozdz2003,drozdz2007,gu2008}. The scaling index $\alpha_r$ for the returns depends also strongly on either $\Delta t$ or $n_{\rm T}$. Typically, one observes a gradual increase of $\alpha_r$ with $\Delta t$ and $n_{\rm T}$ from an initial value of $3 \le \alpha_r \le 4.5$ (see Ref.~\cite{gopikrishnan1999,plerou1999,drozdz2007}), which is usually maintained over a range of the shortest intervals, towards higher values for longer intervals or larger number of aggregated transactions. This increase corresponds to convergence of the distribution of the returns towards a stable, Gaussian distribution.

From pure mathematical perspective, speaking about a convergence is a delicate issue in this context as the convergence towards a Gaussian distribution induced by Central Limit Theorem does not alter the power-law tail slopes. However, the following two remarks have to be made. First, any empirical data is finite and its pdf/cdf tails do not cover the distant regions which may reveal the actual power-law tails that are being repelled towards infinity while CLT exerts its influence. Thus, in econophysics literature it became commonplace that under the notion of "power-law slope", one considers the effective slope of the power function that was best-fitted to empirical distibution in its non-central region. In this study we just follow this approach. Second, one has to keep in mind that financial data is usually non-linearly correlated and cannot be represented by i.i.d. processes~\cite{kwapien2012}. This implies that the standard form of CLT may not be completely adequate in this case, leaving space for its non-extensive generalizations (see Section 3). The exact form of convergence might thus differ from the classic CLT picture.

On the other hand, tail shape of the aggregated trading volume or trade size distributions was a matter of some dispute in literature~\cite{farmer2004}, but nevertheless there is substantial evidence that, for at least the American and Chinese stock markets, the trade size and aggregated volume pdfs can be described by a scale-free tails with the exponent $1.5 \le \alpha_V \le 2.8$~\cite{racz2009,zhou2012,gopikrishnan2000,plerou2007,eisler2007}, i.e., around the L\'evy limit of $\alpha = 2$. This implies that the volume pdf convergence towards a Gaussian distribution is rather slow~\cite{plerou2007}. One of interesting empirical findings is that, for a given market, sometimes a simple relation between both scaling indices exists~\cite{plerou1999,zhou2012,gopikrishnan2000,plerou2007}:
\begin{equation}
{\alpha_r \over \alpha_V} \simeq \xi,
\label{ratio}
\end{equation}
where specific value of $\xi$ depends on a study. For example, Ref.~\cite{gabaix2003,plerou2004,zhou2012} and~\cite{zhou2012} reported $\xi \approx 2$. This relation was found empirically for those data, for which the effective power-law tails were observed. How it behaves for large time scales, for which the distributions converge closer to a Gaussian distribution, has not been studied since the data samples are in such cases rather small. The above relation between the returns and trade sizes or volume (depending on a definition of the returns) can be related to the empirical price impact function $r = f(V)$ describing how a trade or aggregated volume of a given size modifies the price. This function is known to be concave, but its exact form is debatable with possible logarithmic~\cite{potters2003} or (wider documented) power-law relation~\cite{gabaix2003,plerou2004,farmer2004_1,farmer2004,hopman2007}. The latter one can be written as
\begin{equation}
r(V) = c V^{\beta}, \qquad 0 < \beta < 1,
\label{priceimpact}
\end{equation}
where $c$ is some positive constant. For example, Gabaix et al.~\cite{gabaix2003,plerou2004} who considered the American stock market argued that $\beta \approx 1/2$, while Farmer and Lillo~\cite{farmer2004_1,farmer2004} reported $\beta \approx 0.3$ for the London market. It was also seen that this function changes its form under a change of the aggregation time scale, going from strongly nonlinear for small $\Delta t$ or $n_{\rm T}$ to rather linear for larger scales~\cite{bouchaud2009}.

Based on results of their empirical study, Gabaix et al.~\cite{gabaix2003,gabaix2006} formulated a simple theory that was able to explain why both the returns and volume are power-law distributed. They assumed that large market participants - mutual funds, whose activity effectively govern the prices, are power-law distributed and that those funds optimize their trading strategies. Their approach allowed them to derive Eq.(\ref{priceimpact}). This theory opened a debate, in which both the model assumptions and the empirical evidence, which it was grounded on, was criticised~\cite{farmer2004_1,farmer2004,gillemot2006,vaglica2008} and an alternative approach was proposed pointing out to fluctuations of liquidity and not the volume distribution as a principal cause of the form of the return pdf~\cite{bouchaud2009}. However, since the authors of the original model were able to rebut the main objections against it~\cite{plerou2004}, it now seems that the two approaches may be complementary and point out to phenomena that in fact coexist.

Here we do not want to consider empirical data in the light of either theory, but we rather aim at considering statistical distributions of the returns and trading volume for a market which was not analyzed before in this context and compare results with those known from literature for other markets. The Warsaw Stock Exchange (WSE) has total capitalization of about 180 billion USD (August 2013) and, as such, is still counted among the small markets. However, many properties of data from WSE are similar to the respective properties of data from the mature markets~\cite{rak2006,rak2007,rak2008}. Therefore, taking all this into consideration, it is interesting to verify whether any relation similar to Eq.~(\ref{ratio}) can also be identified on WSE.

\section{Results}

We study the distribution of trading volumes and returns based on high-frequency data from the largest 14 companies listed in WSE. Our data cover a time interval starting on Nov 17, 2000 and ending on Mar 6, 2008. The data comprises basically information on all trades which took place in this period but, unfortunately, it does not offer data from the order book. Therefore, we cannot distinguish trades which were buyer- or seller-initiated and have to treat all trades together, which is similar to an approach of Ref.~\cite{gopikrishnan2000,plerou2007}, but different from those in Ref.~\cite{zhou2012,farmer2004}.

\begin{figure}[!ht]
\begin{center}
\includegraphics[scale=.35]{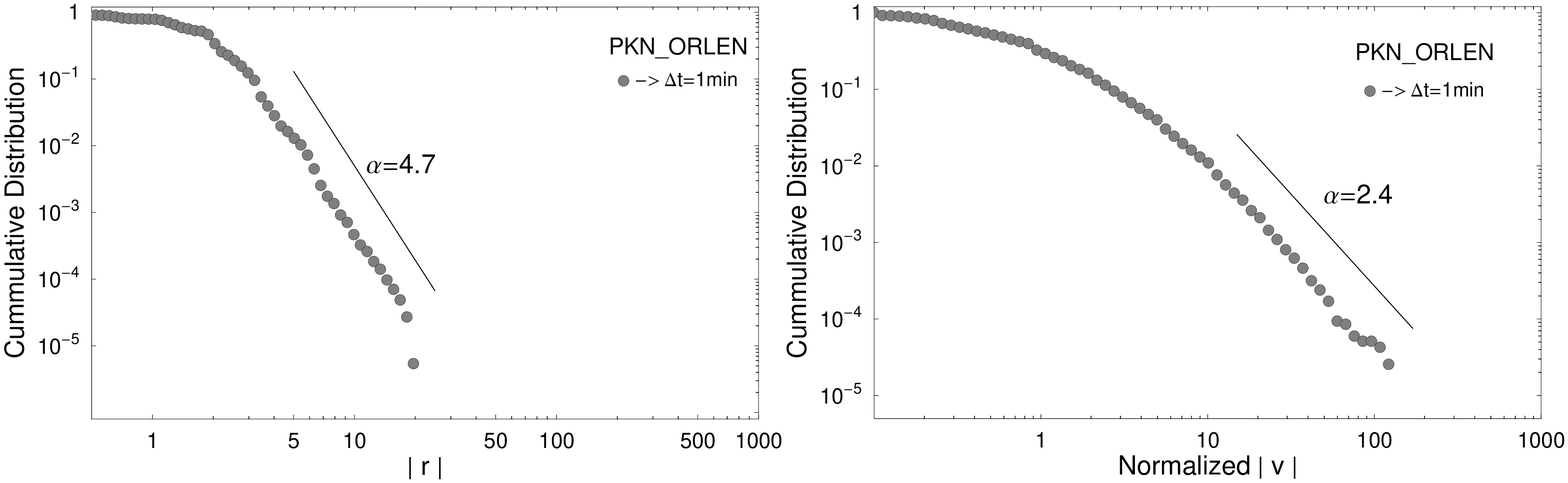}
\end{center}

\vspace{-0.6cm}
\begin{center}
\includegraphics[scale=.35]{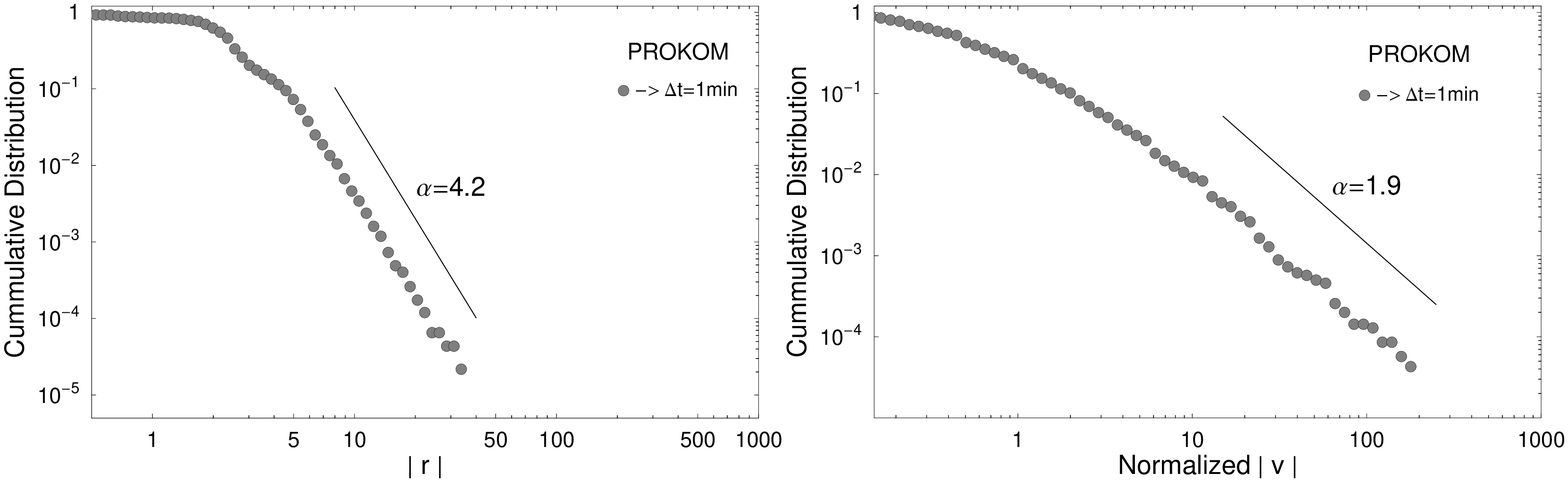}
\end{center}

\vspace{-0.6cm}
\begin{center}
\includegraphics[scale=.35]{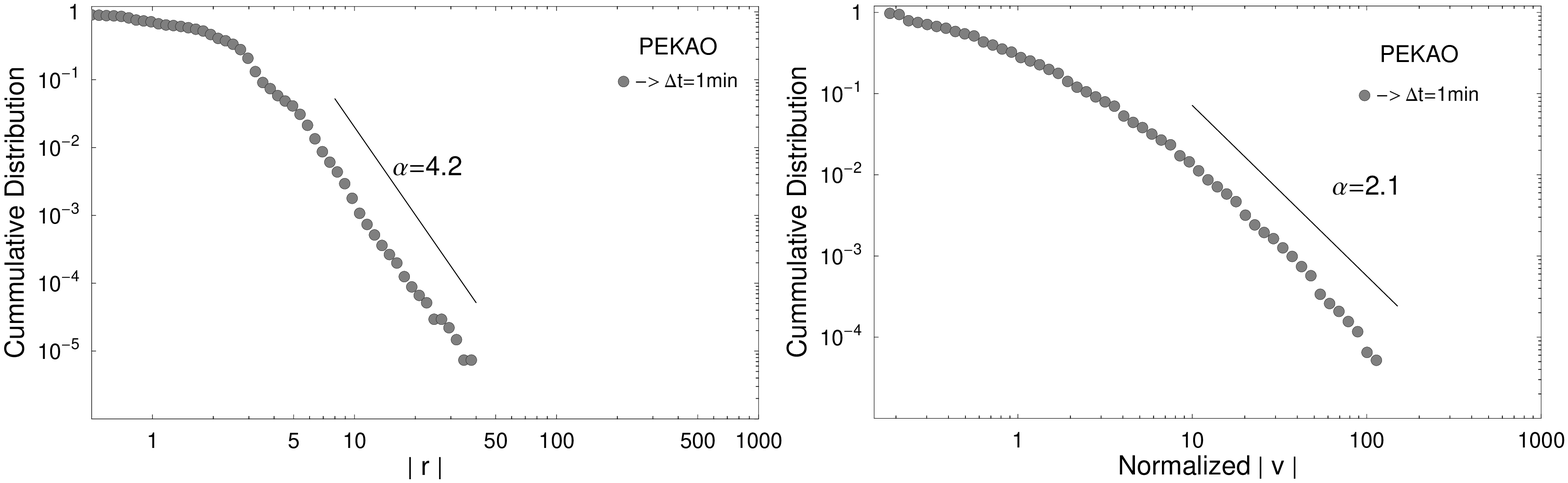}
\end{center}
\vspace{-0.6cm}
\caption{Log-log plot of the cumulative distributions of absolute returns $|r(\Delta t)|$ (left column) and normalized trading volume $V (\Delta t)$ (right column) for $\Delta t = 1$ min. for three Polish companies: PKN Orlen, Prokom, and Pekao over the period Nov 17, 2000 - Mar 6, 2008.}
\label{returns.volume}
\end{figure}

We start our analysis with creating the cumulative distribution functions of the returns and volumes, separately for each company. In this case we consider evenly sampled data with $\Delta t=1$ min. Figure~\ref{returns.volume} exhibits typical results, which indicate that all the return distributions possess power-law tails, while this property is also observed for volume distributions, even though sometimes in a less clear form (like in the case of PEKAO, see the bottom right graph).

In order to compare the scaling exponents of both distributions, we estimated their values by two independent methods: least square fits of power-law functions $f(x)=ax^{-\alpha}$ and the Hill estimators~\cite{hill1975}. For a given time series ${\rm X}$ with its values sorted by their size $X_1 \ge X_2 \ge ... \ge X_N$, the Hill estimator is defined by
\begin{equation}
\texttt{HE}_{\alpha}=\left(\frac{1}{k}\sum_{i=1}^{k}\log(X_{i})-\log(X_{k+1})\right)^{-1}.
\label{he}
\end{equation}
Using both methods for each company, we obtained results that are collected in Table 1. For the returns, the exponents are in the range $3.4 \le \alpha_r \le 5.8$, while for the volumes, the exponents are significantly smaller: $1.6 \le \alpha_V \le 2.4$. It is important that both methods give comparable results. (We also employed the third method of calculating the scaling exponents, namely the Meerschaert-Scheffler estimator~\cite{meerschaert1998}, but it provided us with rather unreasonable estimates, similar for both quantities, so we excluded it from our analysis.)

\begin{center}
\begin{table}[!ht]
\begin{center}
\begin{scriptsize}
\begin{tabular}
{|c|c|c|c|c|c|c|c|c|c|c|c|c|c|}
\hline
\bf {Company name} & $\alpha_r$ & $\alpha_V$ & $\alpha_r / \alpha_V$ & ${\rm HE}_{\alpha_r}$ & ${\rm HE}_{\alpha_V}$ & ${\rm HE}_{\alpha_r} / {\rm HE}_{\alpha_V}$
\\
\hline Agora &4.3&2&2.15&4.04$\pm$0.01&1.64$\pm$0.02&2.46
\\
\hline BRE &4.1&1.9&2.16&4.17$\pm$0.015&1.74$\pm$0.02&2.40
\\
\hline Comarch &4.6&2&2.3&5.00$\pm$0.06&1.85$\pm$0.05&2.70
\\
\hline K\c ety &5.5&2.4&2.3&5.77$\pm$0.04&2.09$\pm$0.03&2.76
\\
\hline KGHM  &4.6&2.4&1.92&5.15$\pm$0.02&2.23$\pm$0.05&2.31
\\
\hline Mostostal Exp. &4.6&2.4&1.92&4.64$\pm$0.01&2.22$\pm$0.01&2.09
\\
\hline Netia &3.4&1.7&2&3.55$\pm$0.02&1.66$\pm$0.01&2.14
\\
\hline Orbis &3.8&1.8&2.11&4.05$\pm$0.01&1.59$\pm$0.03&2.35
\\
\hline Pekao &4.2&2.1&2&5.05$\pm$0.03&1.86$\pm$0.01&2.72
\\
\hline PKN Orlen &4.7&2.4&1.96&5.11$\pm$0.04&2.25$\pm$0.02&2.27
\\
\hline Prokom &4.2&1.9&2.21&4.45$\pm$0.04&1.77$\pm$0.03&2.51
\\
\hline Stalexport &4.7&2.4&1.96&5.12$\pm$0.02&2.16$\pm$0.02&2.37
\\
\hline Softbank &3.9&1.8&2.17&3.73$\pm$0.01&1.76$\pm$0.01&2.12
\\
\hline TP SA &4.6&2.1&2.19&5.13$\pm$0.04&1.68$\pm$0.03&3.05
\\
\hline\hline $\langle \cdot \rangle$ & 4.37& 2.1& 2.09& 4.64& 1.89& 2.45
\\
\hline $\sigma$ & 0.5& 0.26& 0.14& 0.66& 0.25& 0.27
\\
\hline
\end{tabular}
\end{scriptsize}
\end{center}
\vspace{0cm}
\caption{The tail empirical scaling exponents of the cumulative distributions of absolute returns $|r (\Delta t)|$ (second column) and normalized trading volume $V (\Delta t)$ (third column) for 14 Polish companies. The fourth column represents the relation~(\ref{ratio}). The remaining three columns represent the tail exponents and their ratio as given by the Hill estimator~(\cite{hill1975}).}
\end{table}
\end{center}

Interestingly, although the power law relation for the returns and volumes is different than in Ref.~\cite{plerou1999,gopikrishnan2000,plerou2007}, the approximate dependence~(\ref{ratio}) seems to be similar: $\alpha_r / \alpha_V = 2.09 \pm 0.14$ (see the third column in Table 1). The Hill estimator on average gives numbers that are larger than those of power-law fitting (${\rm HE}_{\alpha_r} / {\rm HE}_{\alpha_V} = 2.45 \pm 0.27$) which is rather a typical relation of these two methods (see also~\cite{racz2009}).

\begin{figure}[!ht]
\begin{center}
\includegraphics[scale=.45]{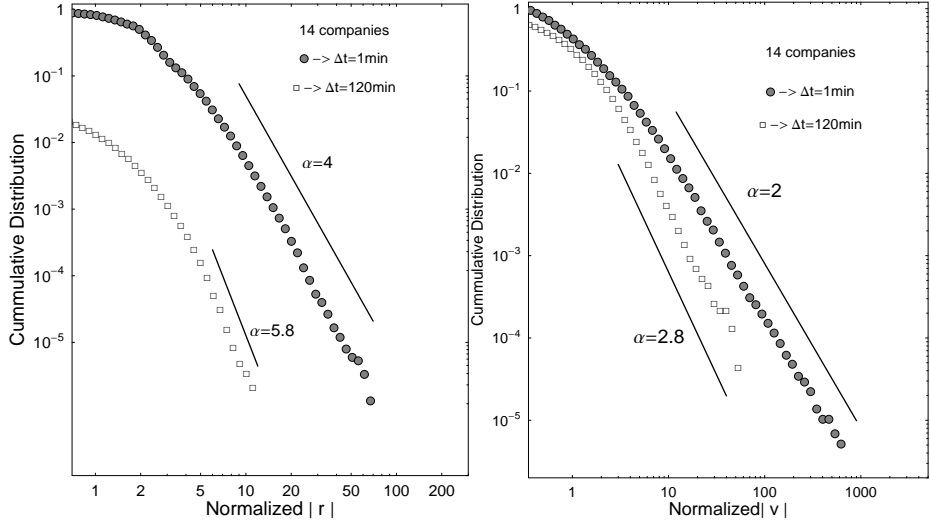}
\end{center}
\vspace{-0.7cm}
\caption{Log-log plot of the cumulative distribution of absolute returns $|r (\Delta t)|$ (left) and of normalized trading volume $V (\Delta t)$ (right) for $\Delta t = 1$ and 120 min over the period Nov 17, 2000 - Mar 6, 2008. The distributions were averaged over the largest 14 companies.}
\label{cdf.comparison}
\end{figure}

To check, whether similar value of $\xi$ holds for different time scales, we constructed analogous time series for longer lags. Figure~\ref{cdf.comparison} shows the cumulative distributions for two lags: $\Delta t=1$ min and $\Delta t=120 $ min for all the companies taken together. It is now clear that, for the Polish stock market, $\xi \approx 2$ holds also for larger values of scaling exponents than $\alpha_r=3$ and $\alpha_V=1.5$ discussed in Ref.~\cite{gabaix2003,plerou2004} in the context of the American markets.

\begin{figure}[!ht]
\begin{center}
\includegraphics[scale=.35]{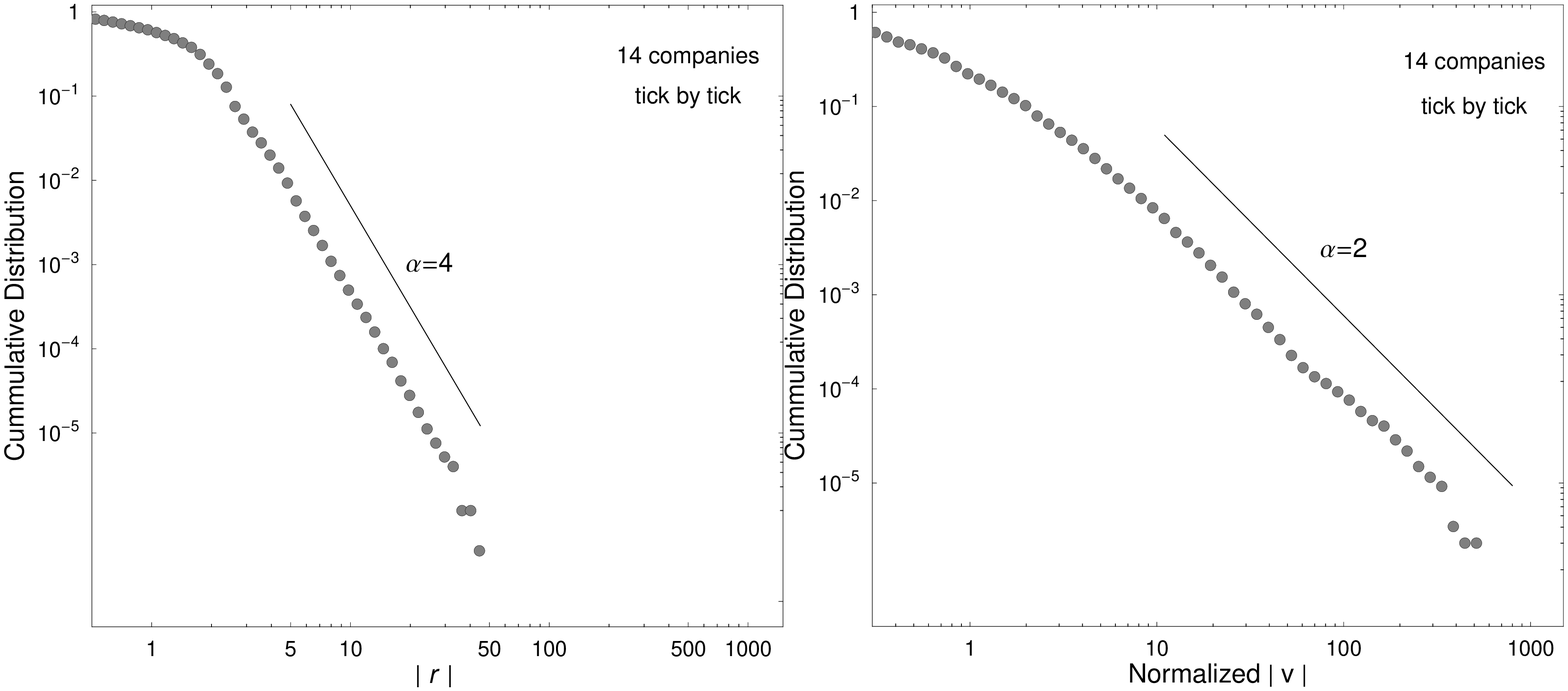}
\end{center}
\vspace{-0.7cm}
\caption{Log-log plot of the cumulative distribution of absolute transaction returns for $n_{\rm T}=1$ (left) and the corresponding normalized transaction volume (right) for the tick-by-tick data over the period Nov 17, 2000 - Mar 6, 2008, averaged over the largest 14 companies.}
\label{tick.by.tick}
\end{figure}

Next, we compare our data with those studied by Zhou~\cite{zhou2012} by fixing $n_{\rm T}=1$. We thus look at time series constructed from the returns caused by individual transactions and the corresponding time series of trade volumes. As we can observe in Figure~\ref{tick.by.tick}, the relation between $\alpha_r$ and $\alpha_V$ is still preserved with $\xi \approx 2$ once again. This outcome is different than that for the Chinese stock market reported in Ref.~\cite{zhou2012}, where the ratio was ca.~1.5, but it is again close to the result for the American stock market~\cite{plerou1999,gopikrishnan2000,plerou2007}.

It can be easily shown that the relation~(\ref{ratio}) is valid always if both the return and the volume distributions have power-law tails and the price impact function $r(V)$ is deterministic with a power-law form as in Eq.~(\ref{priceimpact}). Let $r \sim V^{\beta}$, then we get:
\begin{equation}
x^{-\alpha_r} \sim P(|r| > x) \simeq P(cV^{\beta} > x) = P(V > (xc^{-1})^{1/\beta}) \sim x^{-(1/\beta) \alpha_V},
\label{cdf.indices}
\end{equation}
so
\begin{equation}
\xi \equiv 1/\beta = \alpha_r / \alpha_V.
\label{ratio1}
\end{equation}

\begin{figure}[!ht]
\includegraphics[scale=.35]{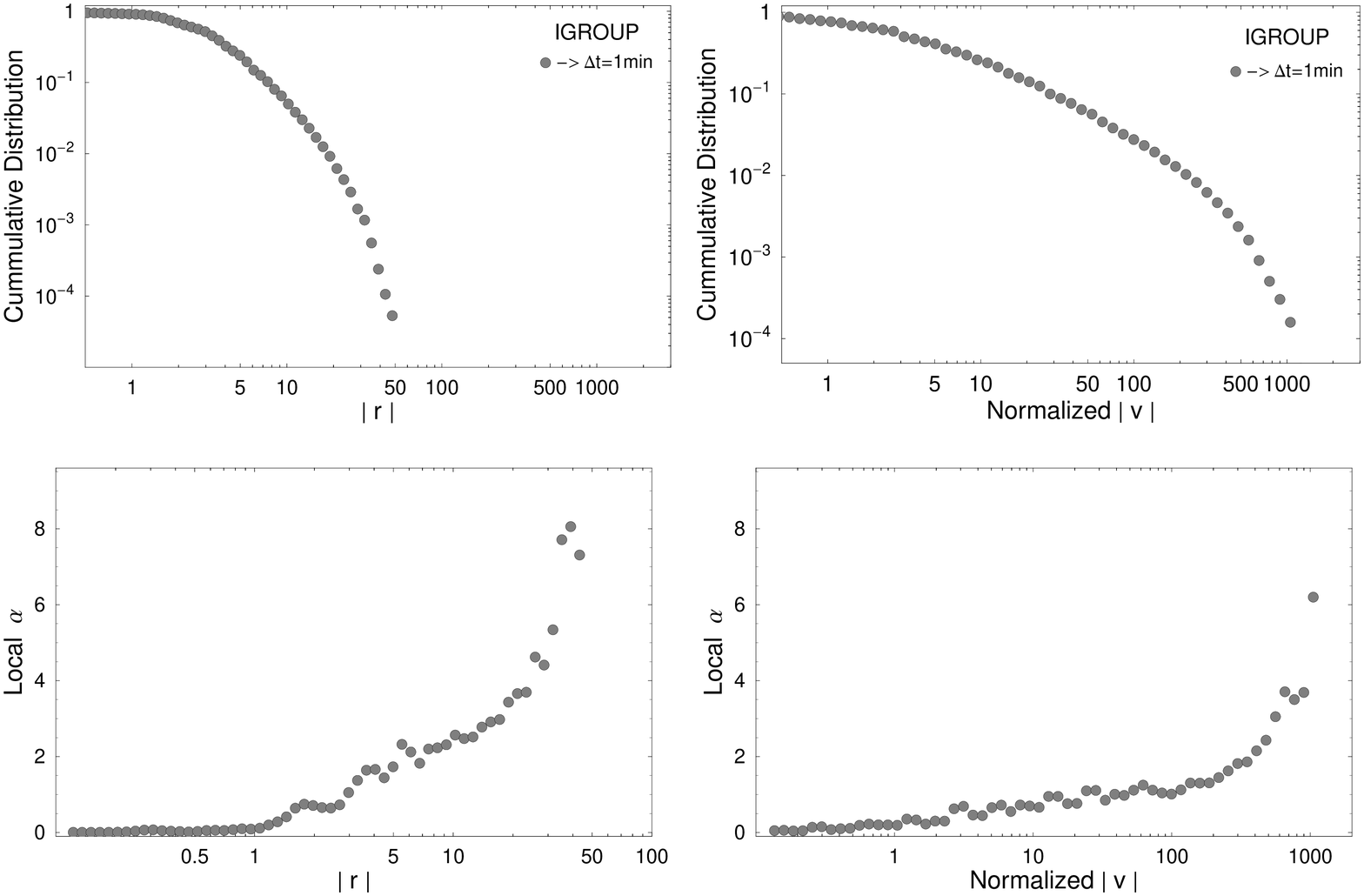}

\begin{center}
\includegraphics[scale=.4]{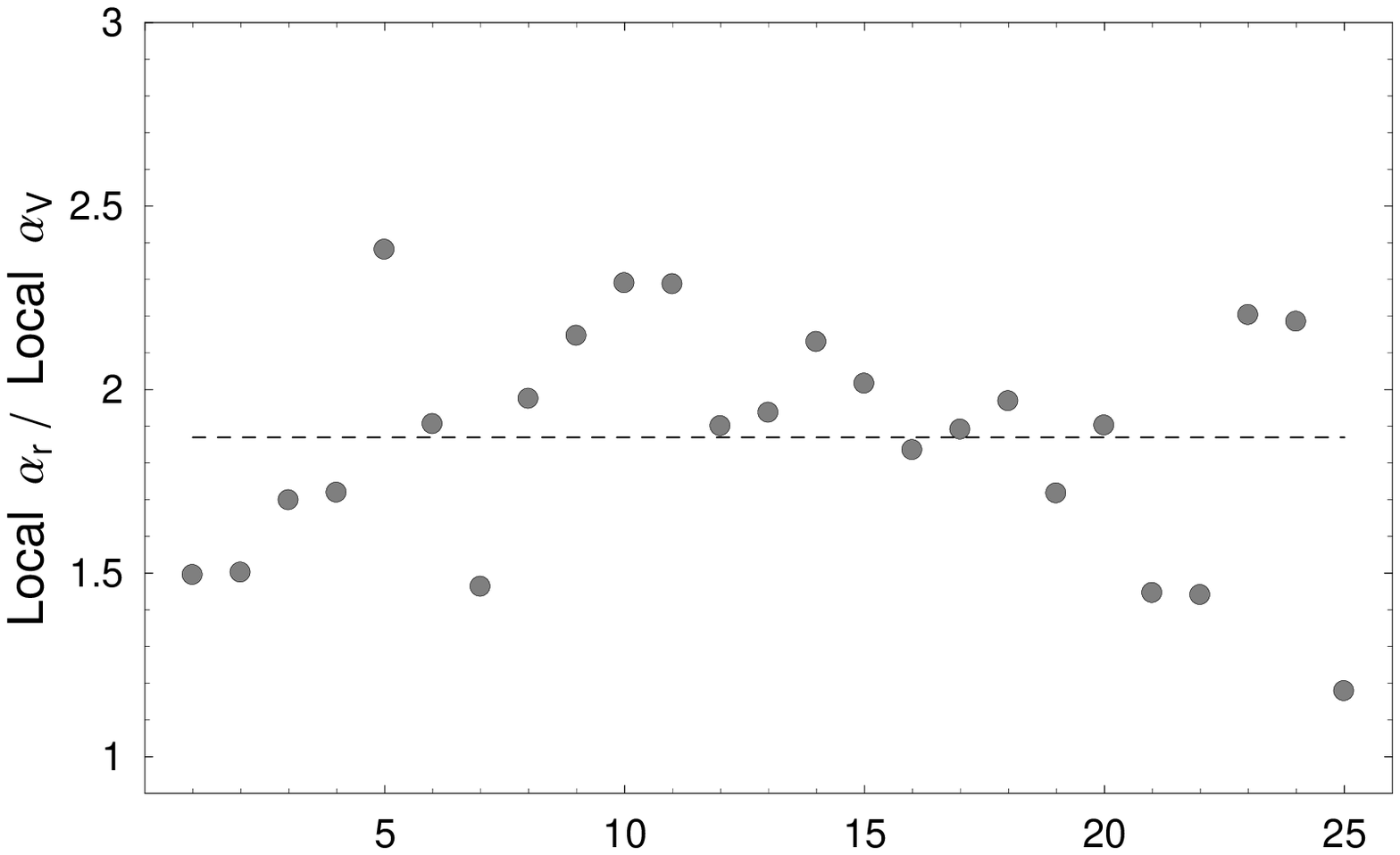}
\end{center}
\vspace{-0.3cm}
\caption{(Top) Log-log plots of the cumulative distributions of absolute returns $r (\Delta t=1 {\rm min})$ (left) and normalized trading volume $V (\Delta t=1 {\rm min})$ (right) for atypical Polish company (IG). (Middle) Locally defined slopes of the distributions shown above. (Bottom) Emprical ratio $\xi = \alpha_r / \alpha_v$ calculated for the corresponding parts of the distributions. Horizontal axis shows the rightmost 25 histogram bins.}
\label{igroup}
\end{figure}

To make our results more complete, we have to mention that in the period considered there was a company (Internet Group) with apparently different dynamics. It developed scale-free tails of the pdfs neither for the returns nor the volume (Figure~\ref{igroup}).

We suspect that this was because it was a highly speculative stock that went through a phase of bankruptcy, followed by a spectacular recovery and then again collapsed towards bankruptcy with no serious institutional investors involved. It is curious, however, that the locally defined slopes of the pdfs are also approximately equal to 2 in this case. Whether this is purely coincident or rather it indicates that the power-law tails are not a necessary condition for the constant ratio of local slopes, we cannot determine at present, but this phenomenon is doubtlessly worth further studies.

\begin{figure}[!ht]
\begin{center}
\includegraphics[scale=0.68]{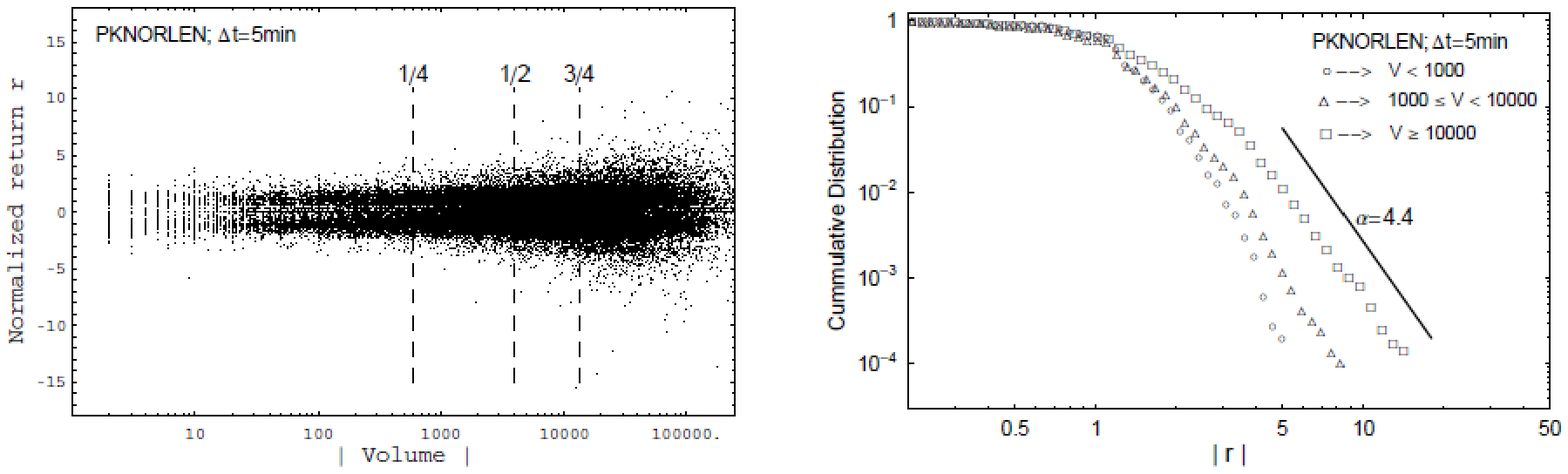}
\end{center}

\vspace{-0.9cm}
\begin{center}
\includegraphics[scale=.68]{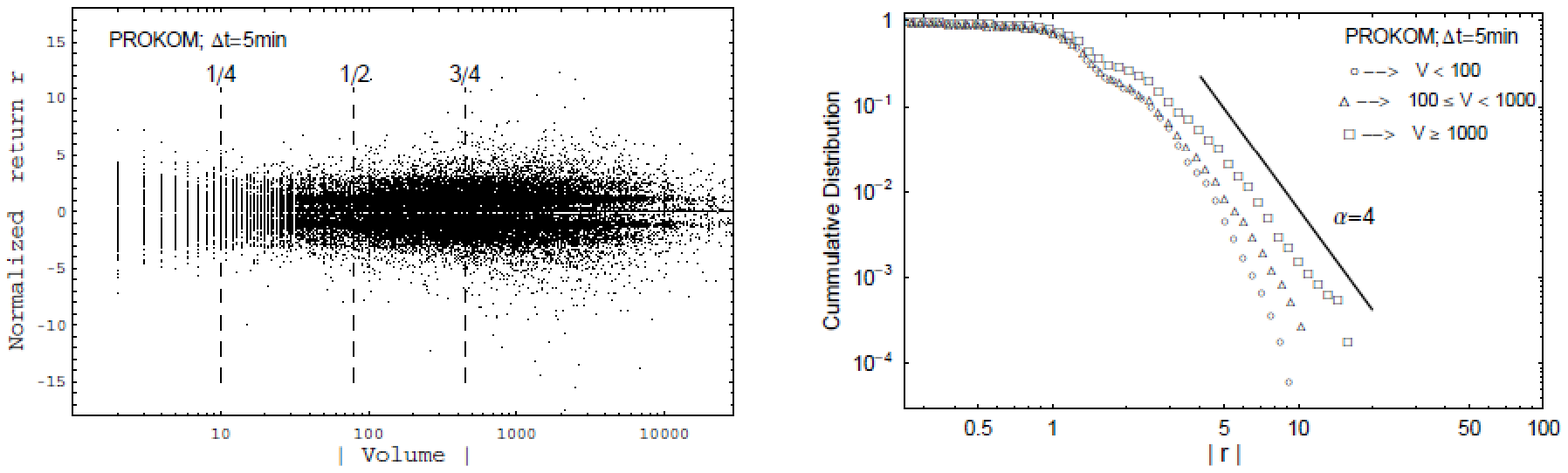}
\end{center}

\vspace{-0.9cm}
\begin{center}
\includegraphics[scale=.68]{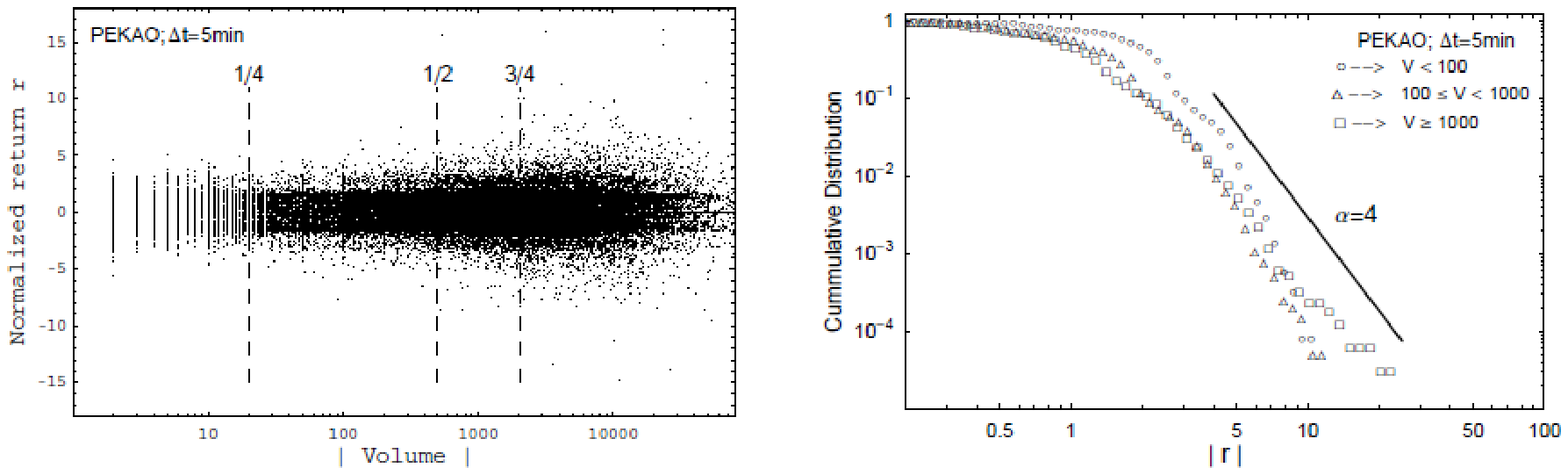}
\end{center}
\vspace{-0.7cm}
\caption{(Left) Absolute returns as a function of volume for three different companies: PKN Orlen, Prokom, Pekao. The vertical dashed lines denote the quartiles of the total number of data points. (Right) The cumulative distributions of absolute returns corresponding to the three volume intervals: circles, triangles, and squares.}
\label{intervals}
\end{figure}

Having shown that, typically, both the return and the volume distribution has scale-free tails, now let us look more closely at the price impact given by Eq.~(\ref{priceimpact}). We address a question if existence of such a price impact can be postulated based on empirical data at our disposal. First, we seek any dependence of the return distribution on the volume size. We create a scatter plot $r$--$V$ for three exemplary companies from different stock sectors (left panels of Figure~\ref{intervals}). By increasing the trading volume, we observe increasing variance of the absolute returns. To better visualize this, we divide the whole volume range into three parts: $V \le 100$, $100 < V \le 1000$ and $V > 1000$ and calculate the return cdfs conditioned on $V$ for each part separately. The resulting distributions are shown in right panels of Figure~\ref{intervals}. For majority of stocks, $\alpha_r$ is noticeably dependent on $V$ in such a way that the larger trading volumes are considered, the gentler is the slope.

Now we can consider the expectation ${\rm E}(r^2|V)$ which was proposed in Ref.~\cite{gabaix2003,plerou2004} as an indication of possible square-root dependence $r(V)$:
\begin{equation}
{\rm E}(r^2|V) = a+bV \ \Rightarrow \ r(V) \sim V^{1/2}.
\label{expectation}
\end{equation}

\begin{figure}[!ht]
\begin{center}
\includegraphics[scale=.40]{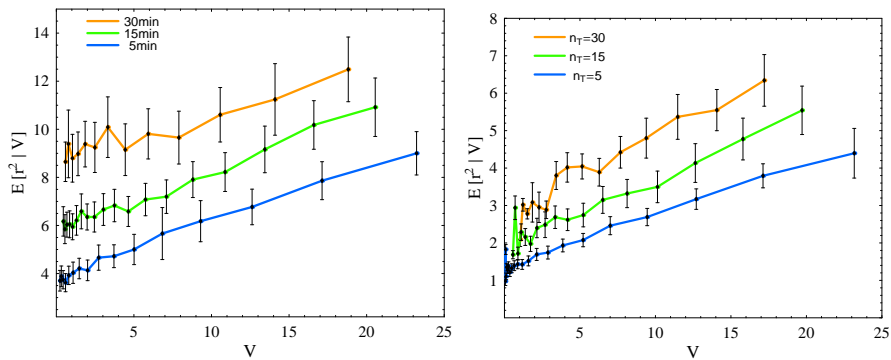}
\end{center}
\vspace{-0.6cm}
\caption{(Left) Conditional expectation ${\rm E}[r^2|V]$ of the squared return $r^2$ (averaged over all 14 companies) for different time scales $\Delta t$, given the aggregated volume $V$.  (Right) The same as on left, but now the volume and returns are aggregated with fixed number of consecutive transactions $n_{\rm T}$ (averaged over all 3 companies:  PKN Orlen, Prokom, and Pekao). The relation ${\rm E}[r^2|V]=a+bV$ is fulfilled for $V \gtrsim 4$.}
\label{expect.original}
\end{figure}

\begin{figure}[!ht]
\begin{center}
\includegraphics[scale=.32]{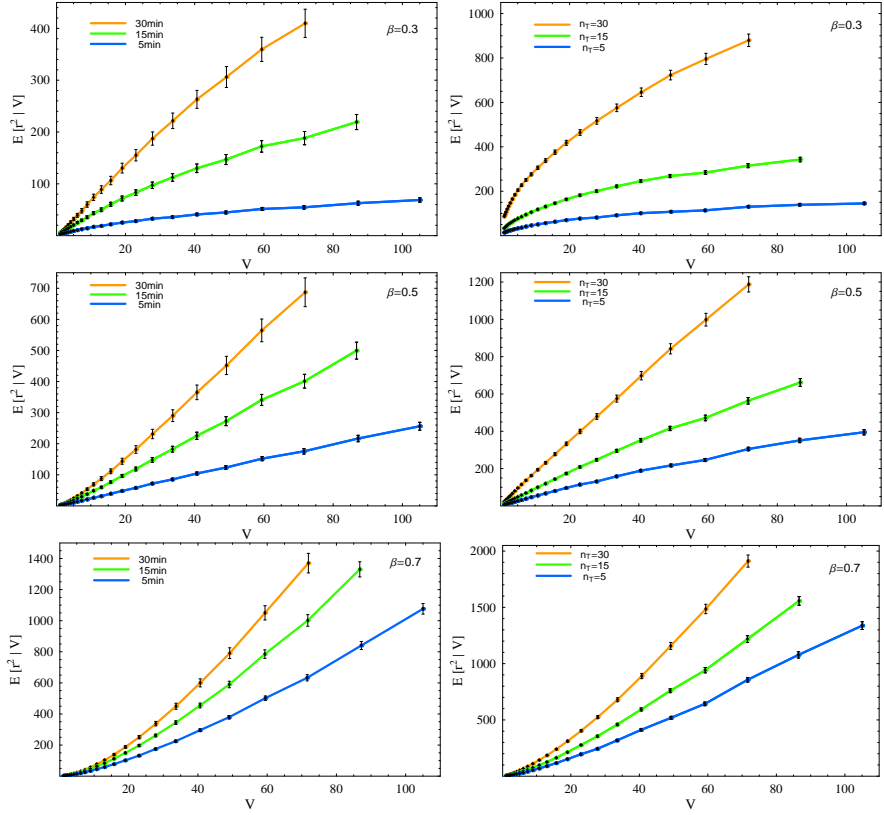}
\end{center}
\vspace{-0.6cm}
\caption{Conditional expectation ${\rm E}[r^2|V]$ for time series of artificial absolute returns created from the original time series of volumes (averaged over all 14 companies) by assuming an exact price impact function $r(V) = cV^{\beta}$ with $c$ being a constant for $\beta=0.3, 05, 07$ respectively. (Left) Volumes $V$ are aggregated over $\Delta t=1$ min consecutive returns. (Right) The same as on left, but here volumes $V$ are aggregated over $n_{\rm T}$ consecutive transactions.}
\label{expect.artificial}
\end{figure}

Since $\xi \approx 2$ suggests that the relation described by Eq.~(\ref{cdf.indices}) holds for the WSE data, we are motivated to look at the empirical form of the price impact function. We compare the expectation ${\rm E}(r^2|V)$ calculated for real (Figure~\ref{expect.original}) and surrogate data (Figure~\ref{expect.artificial}). By following Ref.~\cite{farmer2004_1}, the latter are constructed by assuming the existence of exact price impact function with some exponent $0 < \beta < 1$ in Eq.~(\ref{priceimpact}) and by calculating the artificial returns $r(V(t))$ from the real time series of volumes $V(t)$. If the empirical price impact $r(V)$ is square-root indeed, we expect that ${\rm E}(r^2|V)$ will qualitatively be similar for the actual and the artificial data if we take $\beta=1/2$. On the other hand, after taking $\beta \neq 1/2$, the results in each case have to be different.

In Figure~\ref{expect.artificial}, we show the results obtained for different choices of $\beta$. In agreement with the above arguments, for all the considered $\Delta t$ or for large enough $n_{\rm T}$, the expectation ${\rm E}(r^2|V)$ shows linear behavior for $\beta=1/2$, while for $\beta \neq 1/2$ no obvious linear dependence is seen. These results convince us that the square-root price impact function takes place for Polish stocks, indeed.

\section{$q$-Gaussian fits to volume distributions}

Among the distributions considered in the context of financial fluctuations, there are lognormal distributions, stretched exponentials, truncated L\'evy distributions, and the $q$Gaussian distributions~\cite{rak2007,mantegna1995,borland2002,borland2002a,tsallis2003,malevergne2005}. Our previous research has proved that $q$Gaussians are a good theoretical representation of the empirical return distributions (not only for the stock markets but also for the currency ones)~\cite{drozdz2009,drozdz2010}. Based on this experience, we prefer to fit the financial data with the $q$Gaussians. There are also valid theoretical arguments supporting the use of these distributions in this case~\cite{kwapien2012}.

The $q$Gaussians were discovered in the field of nonextensive statistical mechanics which is a candidate theory to generalize the traditional Maxwell-Boltzmann-Gibbs statistical mechanics to nonequilibrium systems with long-range power-law correlations~\cite{tsallis1988,tsallis1995}. Since from a physicist's point of view it seems that the financial markets can be thought of as systems of this kind, applying the distributions associated with the nonextensive statistics looks natural in this case. The $q$Gaussians are a family of distributions that maximize the nonextensive Tsallis entropy~\cite{tsallis1988} given by:
\begin{equation}
S_q = k_{\rm B} {1 - \int { [p(x)]^q dx } \over q - 1},
\end{equation}
where $p(x)$ is a probability distribution and $k_{\rm B}$ is the Boltzmann constant, under the conditions:
\begin{equation}
\int {x {[p(x)]^q \over \int {[p(x)]^q dx} } dx } = \mu_q, \qquad \int { (x-\mu_q)^2 { [p(x)]^q \over \int {[p(x)]^q dx} } dx } = \sigma_q^2.
\end{equation}
Up to a normalization constant, the formula for $q$Gaussians reads~\cite{tsallis1995,tsallis1998}:
\begin{equation}
G_q(x) \sim \exp_q [-\mathcal{B}_q(x - \mu_q)^2 ],
\end{equation}
where $\exp_q x = [1 + (1 - q) x]^{1 \over 1 - q}$ and $\mathcal{B}_q = [(3 - q) \sigma_q^2]^{-1}$. The $q$Gaussians are defined for $0 < q < 3$. Unlike e.g. the Pareto and other already-listed distributions, the $q$Gaussians can consistently fit the whole range of fluctuations, not only the tails. The asymptotic behaviour of $q$Gaussians is of the power-law type with the scaling exponent $\alpha\equiv\alpha_{qG}$ uniquely determined by $q$ according to the relation:

\begin{equation}
\alpha_{qG} = {3 - q \over q - 1}.
\label{qG}
\end{equation}
In particular, $q \ge 5/3$ corresponds to $\alpha_{qG} \le 2$, i.e. to the L\'evy-stable regime.

\begin{figure}[!ht]
\includegraphics[scale=.45]{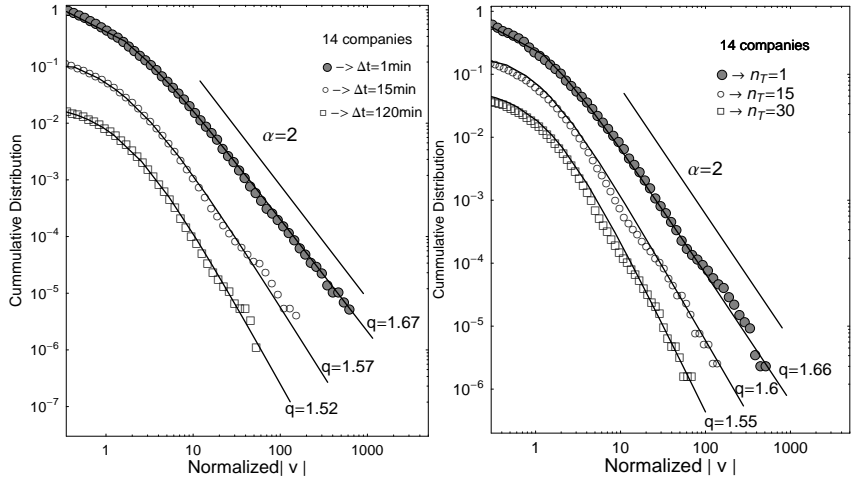}
\vspace{-0.7 cm}
\caption{Log-log plot of the cumulative distributions of the aggregated trading volume for different time scales $\Delta t = 1$, 15, and 120 min (left), as well as for different number of consecutive transactions $n_{\rm T}=1$, 15, and 30 (right). The solid lines represent the best theoretical fits of the cumulative $q$Gaussians~\cite{rak2007} with appropriate $q$'s.}
\label{volume.qgaussians}
\end{figure}

We consider trading volume distributions for different stocks listed in WSE. For each stock, we compare the cumulative distributions obtained for different time scales $\Delta t$ or different number of aggregated trades $n_{\rm T}$. All the cumulative distributions are fitted by appropriate $q$Gaussian cdfs~\cite{rak2007}. Exemplary results are shown in Figure~\ref{volume.qgaussians}. The agreement between the empirical data and the fits is encouraging: for all the time lags considered in our study (shown and not shown), one obtains a good theoretical representation of the data over the whole range of values. In each case, the inaccuracy does not exceed a few largest events. The $q$Gaussian fits to the tick-by-tick data are also quite good. Moreover, all values of the empirical scaling exponents ($\alpha_V$) are similar to values of $\alpha_{V_{qG}}$ for appropriate $q$'s.

\section{Summary}

In the present contribution, we focused our attention on a relationship between large returns and trading volume based on data from Warsaw Stock Exchange, which is an emerging market. Our intention was to investigate whether any systematic relation between the distributions of returns and volumes exists in this case, similar to outcomes of earlier works which were focused on larger markets like the American, Chinese, London, and Paris ones.

We have shown that the relation~(\ref{ratio}) with $\xi \approx 2$ (or $\approx 2.5$ based on the Hill estimator) holds for majority of the analyzed WSE stocks. We also observed that the price impact of trading volume can be modelled by a square-root function~(as described by Eq.(\ref{priceimpact}) with $\beta=1/2$). These outcomes go in parallel with some earlier studies (see especially Ref.~\cite{plerou1999,gabaix2003,plerou2004,gopikrishnan2000,plerou2007}). Another curious result of our study worth further investigation is that one can observe a relation between the local slopes of both types of the distributions even if they do not display any power-law tails. However, this result was obtained only for a single company and therefore we do not want to draw any decisive conclusions based on it yet.

An additional interesting observation done in our work is that not only the distributions of returns but also the distributions of trading volumes can be well described by the $q$Gaussian functions.

\end{document}